\documentclass[12pt, a4paper]{article}
\usepackage{graphicx}
\title{Multiple time scales and the empirical models for stochastic volatility}
\author{G.L. Buchbinder\footnote{Corresponding author. \newline\indent \textit{E-mail:} glb@omsu.ru},
K.M. Chistilin \\\small{\textit{ Department of Physics, Omsk State
University, 55a Peace Avenue,}} \\\small{\textit{ Omsk 644077,
Russian Federation}}}
\date{}
\begin{document}
\maketitle

\section*{\large Abstract}
The most common stochastic volatility models such as the
Ornstein-Uhlenbeck (OU), the Heston, the exponential OU (ExpOU)
and Hull-White models define  volatility as a Markovian process.
In this work we check of the applicability of the Markovian
approximation  at separate times scales and will try to answer the
question which of the stochastic volatility models indicated above
is the most realistic. To this end we consider the volatility at
both short (a few days) and long (a few months)time scales as a
Markovian process and estimate for it the coefficients of the
Kramers-Moyal expansion using the data for Dow-Jones Index. It has
been found that the empirical data allow to take only the first
two coefficients of expansion to be non zero that define form of
the volatility stochastic differential equation of It\^o. It
proved to be that for the long time scale the empirical data
support the ExpOU model. At the short time scale the empirical
model coincides with ExpOU model for the small
volatility quantities only.\\
{\small \textit{PACS:} 89.65.Gh; 02.50.Ey; 02.50.Ga; 02.50.Cw\\
\noindent\textit{Keywords:} Stochastic volatility models;
Volatility autocorrelation; Leverage; Fokker-Plank equation}
\newpage
\section*{\large 1. Introduction}
The stochastic volatility (SV) models with continuous time have
been introduced into  literature in late  of 80-s of the last
century \cite{S87,SS91,H93,HW87}. According to these models the
market dynamics is the two-dimensional stochastic process in which
the asset price $S_t$ obeys the stochastic differential equation
in the It\^{o} form ( the index $t$ is omitted for simplicity)
\begin{equation}
dS = \mu S dt + \sigma  S dW_1(t) ,\label{1}
\end{equation}
where the parameter $\mu$ is the drift coefficient, $W_1(t)$ is a
standard Wiener process  and $\sigma_t $ is the volatility
considered as a stochastic variable.

The empirical analysis have established two important stylized
facts concerning with the volatility. Firstly, this process has a
long memory emerging, in particular, in that the autocorrelation
function absolute returns decay very slowly with time. One can
separate, at least, two  characteristic time scales in the
behavior of the autocorrelation function. At the initial stage
there is a fast decay on the short time scale of the order of few
days followed by the slow decay during a few months, defining the
long time scale. Secondly, there is the negative correlation
between past returns change and future volatility (so-called
"leverage" effect).

At the present   different SV models are discussed in literature.
To a certain extent these models are based either on the model of
the geometrical Brownian motion or originate from
Ornstein-Uhlenbeck (OU) process. It is assumed that  volatility is
a function $\sigma = g(Y)$ of a stochastic process $Y_t$ and that
the dynamic equation for $Y_t$ can be represented as a stochastic
differential equation in It\^o form
\begin{eqnarray}
&&dY = \alpha (m - Y)dt +  f(Y)dW_2(t) ,\label{2} \\
&&\sigma = g(Y)\, .\nonumber
\end{eqnarray}
Eq.(\ref{2}) defines the so-called class of mean-reverting
processes in which $Y_t$  goes to mean value $m$ at $t\to \infty$
with the velocity $\alpha$.The quantity $1/\alpha$ is the time of
relaxation of $\sigma$ to its  equilibrium value  approximately
equal to $g(m)$ and actually represents the characteristic time
scale of the process. The Wiener process $W_2(t)$  in general is
correlated with the process $W_1(t)$.

Depending on $g$ and $f$ one can distinguish basically four
frequently used SV models.\\
1) The Ornstein-Uhlenbeck (OU) model \cite{S87,SS91} with $\sigma
= Y$ and $f(y)= k$, where $k$ is a positive constant and

\begin{equation}
d\sigma = \sigma (m- Y)dt + k dW_2(t) \label{3}
\end{equation}
2) The exponential Ornstein-Uhlenbeck (ExpOU) model \cite{S87}
with $\sigma = e ^{y}$ and $f(y)= k > 0$, where it is assumed that
$Y = \ln\sigma$ follows to the OU process and variable $\sigma$,
as it is easy to show in this case, satisfies the equation

\begin{equation}
d\sigma  = \sigma \left[
\frac{k^2}{2}-\alpha(\ln\sigma-m)\right]dt + k \sigma dW_2(t)
\label{4}
\end{equation}
3)The Heston model \cite{H93} where $\sigma = \sqrt{y}$ and $f(y)
= k\sqrt{y}$, $(k > 0)$. In this model  it is assumed that the
volatility is the OU process of the form (\ref{3}).

\noindent 4) The Hull-White model \cite{HW87}with $\sigma =
\sqrt{y}$, $f(y) = ky$, $(k > 0 ))$ and

\begin{equation}
d\sigma  = \frac{1}{2\sigma}\left[ \alpha(m- \sigma^2)-
\frac{k^{2}\sigma^{2}}{4}\right]dt + \frac{k\sigma}{2} dW_2(t)
\label{5}
\end{equation}

Originally the parameters of the models 1) - 4) were being
estimated by a fitting to the  empirical  data from options
pricing. Lately the  question about the applicability of one or
the other stochastic volatility model for describing  a time
evolution of stock prices, market indices or exchange rates is
actively discussed in the physical literature. The fundamental
problem is  finding the most realistic  model and estimating its
parameters.

One of the approaches is that parameters of a model are estimated
by  fitting the theoretical probability distribution functions
(PDF) of returns to the empirical curves. So the studies carried
out in works \cite{DY02,SY03,SP04,BC05} have showed that the
Heston model well enough reproduce  the empirical distributions
for Dow - Jones Index and a number of stocks. On the other hand,
in the case of the high-frequency data,  the Heston model, as well
as Hull-White model, applied  to German DAX Index give the return
distributions not conforming to tails of the empirical curves
\cite{RM04}(see also \cite{SM02}).

The alternative approach is to estimate the parameters of the
above models in such a way as to reproduce  other the market
stylized facts.In particular, in  works
\cite{PM03,MP02,PMA04,PM05} the parameters of the OU, Heston and
ExpOU models have been estimated by comparison of the theoretical
predictions to the observed leverage effect. It has been showed
that these models qualitatively reproduce the observed effect as a
result of the choice of the parameters, however, the empirical
data do not allow to assert the most appropriate  model. As
regards the autocorrelation function, in contrast with other
models, ExpOU qualitatively reproduces the behavior of the
empirical curve at medium and long times by  fitting the
parameters. On the other hand in order to take into account the
occurrence of two time scale in the work \cite{PMB04} (see also
\cite{V06}) a three dimensional diffusion model, assuming that the
mean reverting level $m$ is random, has been introduced.

Thus  by fitting the parameters above SV models sometimes can
reproduce well enough  the probability densities of the returns or
describe specific  observed authoritativeness (the leverage
effect, behavior of autocorrelation function). However the
question of choice of  most realistic  stochastic volatility model
still remains open.

Models 1) - 4) determinate the volatility as a Markovian process.
This follows from the well-known  fact that  solutions of a SDE of
It\^o have the Markovian properties \cite{G97}. At the same time
the empirical volatility autocorrelation function, decaying very
slowly with time, shows , in general, the non-Markovian behavior.
Furthermore as it has been noted above the  autocorrelation
function has at least two characteristic time scales. In this
connection it is worth noting that the empirical analysis reveals
the presence of a well-separated time scales in the dynamics of
the volatility itself.

So, LeBaron in \cite{B01} has showed that the SV model, where the
volatility behavior at short ($\sim 1$ day), medium ($\sim 5$
weeks) and long ($\sim 5$ years) time scales  is defined by three
different stochastic processes, reproduces power law in the
asymptotic  of log returns of the Dow-Jones Index and long memory
in the volatility fluctuations.

In the work  of J.-P. Fouque at al \cite{F03} (see also \cite{F3})
volatility dynamics both at short (a few days) and long (few a
months) time scales was considered within the scope of the ExpOU
model but with  different relaxation times for  each scale. For
S\&P 500 high-frequency data the short time scale has been found
of order $\alpha ^{-1} \sim 1.5$ days.

In this work  we want to check the application of the above SV
models separately both at short and long the time scales.  If at
the specific  time scale the Markovian approximation is
applicable, then the coefficients of the SDE of It\^o written as

\begin{equation}
d\sigma = D_1(\sigma)dt + \sqrt{2D_2(\sigma)}dW_2(t) \label{6}
\end{equation}
can be obtained from the known expression of the theory of
Markovian processes \cite{R84}

\begin{equation}
D_k (x) = \mathop {\lim }\limits_{\tau \to 0} \frac{1
}{\tau}\frac{1 }{k!}\int\ {dx'(x' - x)^kp(x',t + \tau \vert x,t )}
, \label{7}
\end{equation}
where $p(x',t'|x,t)$ is the conditional PDF  and $D_k$ are the
coefficients of the Kramers-Moyal expansion. Such approach has
recently been used in \cite{FPR0,RPF01} and allowed directly from
the data to estimate the coefficients of SDE of It\^o for returns
handling high-frequency dynamics of DEM/USD exchange rates. Here
this method is applied to both high-frequency and low-frequency
data. The SDEs of the form (\ref{6}) obtained for this two data
sets have to define the volatility behavior both at short and long
time scales. In the end this gives an opportunity to make a
comparison with the known  SV models and, to a certain extent, to
answer the question of how consistent one or the other model is
with the empirical data.

The paper is organized as follows. Section 2 is devoted to
describing  the method of determination of the volatility time
series. In Section 3 the coefficients of the Kramers-Moyal
expansion  have been obtained for  different time scales.   In
Section 4. the numerical solution of the Fokker-Plank equation for
conditional PDF is given and the convergence of the solutions  to
the stationary distributions is considered  for both time scales.
In section 5 on the basis of the obtained SDE  for volatility the
simulation of the return   series is carried out and its
properties are studied.  The analysis of the results is given in
conclusion.

\section*{\large 2. The estimation of volatility and data sets}
Unlike prices changes the volatility is not directly observed. At
the present there are a different methods  of its estimation (for
example, see \cite{PM05,Ya99}). Most frequently the volatility at
the moment of time $t$ is estimated as the standard deviation
\begin{eqnarray}
&&\sigma_t^2=\frac{1}{T}\sum\limits_{t'=t}^{t+T}(r_{t'}-<{r_{t}}>)^2,
\label{8}\\
&&<r_{t}>=\frac{1}{N}\sum\limits_{t'=t}^{t+T}r_{t'},\nonumber
\end{eqnarray}
where $r_{t}=\ln S_t/S_{t-\Delta}$ are log-returns and an average
is carried out over time window $T=(N-1)\Delta$ with an integer
$N$.

The two different data sets have been used for the empirical
analysis: the high-frequently data set (HFD) for the Dow-Jones
Index (data sampled at 5 min intervals from Feb. 16, 2001 to Feb.
26, 2005\footnote{http://www.finam.ru}) and low-frequently data
(LFD) (with daily data for the Dow-Jones Index from Jan. 2, 1990
to Feb. 25, 2005\footnote{http://finance.yahoo.com }).  One has
respectively $\Delta=5$ min, T=2 hours  for HFD and $\Delta=1$
day, T=1 month (21 days) for LFD. The non-overlapping intervals of
averaging T have been used for calculation of the volatility given
Eq.(\ref{8}) and respectively the sampling time interval for the
volatility data  equals T.

The obtained empirical values $\sigma_t$ have been used for the
construction of the stationary distributions of the volatility and
the conditionals PDFs. In doing so it has been assumed that the
volatility is a stationary process \cite{PM05}.

\section*{\large 3. The estimation of the Kramers-Moyal
coefficients}

According to what was said in the introduction we shall consider
the volatility on the above indicated time scales in the Markovian
approximation. In this case, as it is known, the conditional
probability density obeys a master equation in the form of a
Kramers-Moyal expansion \cite{R84}.

\begin{equation}
 \frac{\partial }{\partial t }p(\sigma,t \vert \sigma_0 ,t _0 ) =
\sum\limits_{k = 1}^\infty {\left( { - \frac{\partial }{\partial
\sigma}} \right)^kD_k (\sigma)p(\sigma,t \vert \sigma_0 ,t _0 )}
\label{9}
\end{equation}
where the coefficients $D_k$ are defined as
\begin{equation}
D_k (\sigma ) = \mathop {\lim }\limits_{\Delta t \to 0} \frac{M_k
(\sigma,\Delta t )}{\Delta t }\label{10}
\end{equation}
and moments $M_k$ are
\begin{equation}
M_k (\sigma,\Delta t ) = \frac{1 }{k!}\int\ {(\tilde {\sigma} -
\sigma)^kp(\tilde {\sigma},t + \Delta t \vert \sigma,t )d\tilde
{\sigma}}\label{11}
\end{equation}

In this section we shall calculate the coefficients $D_1$ and
$D_2$ of the expansion (\ref{9}) and show that with enough
accuracy the data set allows to take $D_4$ to be zero. According
to Pawla's theorem \cite{R84} at $D_4=0$  all coefficients $D_k$
with $k\geq 3$ vanish and the equation (\ref{9}) reduces to a
Fokker-Plank equation. In this case it is coefficients $D_1$ and
$D_2$ that define the form of SDE for volatility of the form
(\ref{6}).

For the calculation of the moments $M_k$ the conditional densities
$p(\tilde {\sigma},t + \Delta t \vert \sigma,t )$ (see Fig.4) have
been determinated from the empirical data and the numerical
integration in (11) has been performed. Further the approximation
of the limiting passage $\Delta t \rightarrow 0$ in (10) has been
employed and coefficients $D_1$ and $D_2$ have been obtained.

Fig.1 shows some typical dependence of the moments $M_{1,2}$ of
$\Delta t$. In order to obtain the moments $M_{1,2}(\sigma,\Delta
t)$ for the case of the small $\Delta t$, the volatility given
Eq.(\ref{8}) has been calculated  using  the overlapping intervals
T. At the small $\sigma$ the moments $M_{1,2}$ are well enough
described by the linear dependence on $\Delta t$. Therefore the
limit in (\ref{10}) has been approximated as follows
\begin{equation}
\mathop {\lim }\limits_{\Delta t \to 0} \frac{M_k (\sigma,\Delta t
)}{\Delta t } \approx \frac{M_k (\sigma,\Delta t )}{\Delta t
}\label{12}
\end{equation}
at $\Delta t = T$. At large $\sigma$ values of moments fluctuate
drastically because of the decrease of the statistical data.
Nevertheless here too limit  (\ref{10}) has been approximated by
relationship (\ref{12}) at $\Delta t = T$.

The results  of the calculation of the coefficients
$D_{1,2}(\sigma)$ given by Eqs.(\ref{10}-\ref{11}) are shown in
Fig.2. It has turned out that for both data sets $D_1(\sigma)$ can
 be approximated well enough by the function that coincides in form
with the drift coefficient of the ExpOU model (\ref{4}) (Fig.2a;
2c). For approximation of the coefficient $D_2(\sigma)$ for LFD
the square dependence on $\sigma$ has been used (Fig. 2d). For HFD
for the small $\sigma$, $D_2(\sigma)$ can also be approximated by
the square dependence on $\sigma$, however, for large $\sigma$ it
increases faster then the square function. Therefor for
approximation $D_2(\sigma)$ for all $\sigma$ the function has been
used
\begin{equation}
D_2(\sigma)=b_1\sigma^2\exp(b_2\sigma).\label{13}
\end{equation}
In the result of the fitting we have obtained:

\noindent for HFD
\begin{eqnarray}
&&D_1^{(H)}(\sigma)=-\sigma (a_1 - a_2
\ln(\sigma/\sigma_0))\label{14}\\
&&D_2^{(H)}(\sigma)=b_1\sigma^2\exp(b_2\sigma)\label{15}
\end{eqnarray}
where $a_1=-0.071\, {\rm (month)}^{-1}$; $a_2=26.5\, {\rm
(month)}^{-1}$; $b_1=7.08\, {\rm (month)}^{-1}$; $b_2=2.65\, {\rm
(month)}^{1/2}$;
\noindent for LFD
\begin{eqnarray}
&&D_1^{(L)}(\sigma)=-\sigma (a_1 - a_2
\ln(\sigma/\sigma_0))\label{16} \\
&&D_2^{(L)}(\sigma)=b_1\sigma^{2}\label{17}
\end{eqnarray}
where $a_1=4.47\,{\rm (month)}^{-1}$; $a_2=0.41\,{\rm
(month)}^{-1}$; $b_1=0.06\,{\rm (month)}^{-1}$. For both data sets
$\sigma_0$ is the mean volatility equal to $0.044\,
{\rm(month)}^{-1/2}$. Correspondingly, the relaxation times  are
$1/\alpha^{(H)}=1/a_2^{(H)}=0.79$ day and
$1/\alpha^{(L)}=1/a_2^{(L)}=2.43$ month.

At last, the results of the calculation of the coefficients $D_4$
are represented in Fig.3. It is shown  that the values of the
coefficient $D_4$ in fact are equal to zero, the fluctuations do
not exceed $10^{-2}$ for HFD and $3\cdot10^{-7}$ for LFD which is
a few orders less than the corresponding values of $D_1$ and
$D_2$.

\section*{\large 4. The numerical solution of the Fokker-Plank
equation}

As it has been noted in the previous section at $D_4=0$ the master
equation (\ref{9}) reduces to the Fokker-Plank equation

\begin{equation}
 \frac{\partial }{\partial t }p(\sigma,t_0\vert\sigma_0,t_0) = \left\{ { -
\frac{\partial }{\partial \sigma}D_1 (\sigma) + \frac{\partial
^2}{\partial \sigma^2}D_2 (\sigma)}
\right\}p(\sigma,t_0\vert\sigma_0,t_0) \label{18}
\end{equation}

In this section we shall consider on the basis of a numerical
solution of Eq.(\ref{18}) the time evolution of the conditional
densities $p(\sigma,t|\sigma_0,t_0)$ and show that the stationary
solution of this equation is   consistent enough with the
empirical densities.

Eq.(\ref{18}) was solved with at boundary conditions at $\sigma=0$
and $\sigma=\infty$

\[p(\sigma,t_0\vert\sigma_0,t_0)|_{\sigma=0;+\infty}=0\hspace{1cm} (\sigma\geq 0)\]
and the initial condition
$p(\sigma,t_0\vert\sigma_0,t_0)=\delta(\sigma-\sigma_0)$ where
$\delta(x)$ is $\delta$-function and $\sigma_0=0.044$. For the
numerical solution of the Fokker-Plank equation (\ref{18}) the
finite-difference method given in [22] has been used.

The conditional densities $p(\sigma,t|\sigma_0,t_0)$ for different
times $t$ are represented in Fig.4. As it is shown from Fig.4a for
high-frequency data the stationary state is reached within
$t\simeq1.5$ days (from this on time the theoretical curves
practically coincide). For low-frequency data the stationary state
is settling within the time of the order of 5 months (Fig.4c). As
it is shown from Fig.4b;4c the theoretical stationary
distributions are consistent enough with the empirical volatility
densities. To some extent this fact can serve as validation of
estimating the coefficients $D_{1,2}$.

\section*{\large 5. The simulation  of the return series}

As it is known \cite{G97,R84} the Fokker-Plank equation (\ref{18})
is equivalent to the SDE of It\^o  of the form
\begin{equation}
d\sigma = D_1(\sigma)dt + \sqrt{2D_2(\sigma)}dW_2(t) \label{20}
\end{equation}
This equation in combination with Eq.(\ref{1}) enables to perform
the simulation of the prices series that gives an opportunity to
obtain a theoretical return PDF.

In order to eliminate the parameter $\mu$ in Eq.(\ref{1}) let us
introduce the new variable $x_t=\ln S_t/S_0-\mu t$, where $S_0$ is
the initial price. It is easy to obtain that
\begin{equation}
dx_t = -\frac{\sigma_t^{2}}{2}dt + \sigma_tdW_1(t). \label{21}
\end{equation}

Using Eqs.(\ref{20}) and (\ref{21}) and the explicit form of the
coefficients $D_1$ and $D_2$ for both data sets we have generated
the series $x_t$. The Wiener processes $W_1(t)$ and $W_2(t)$ have
been assumed to be independent. The found price series have been
used for the  plotting of the probability density and the
autocorrelation function of the absolute log-returns.

Fig.5 represents the plots of PDF of the prices changes $\triangle
x=\ln S_t/S_{t-\triangle}-\mu\triangle$ obtained from both the
generated price series and the empirical data. As it is seen there
is a good agreement between the corresponding curves.

The plots of the autocorrelation function of the absolute
log-returns $|r_t|$ are given in Fig.6. In the case of
high-frequency dynamics (Fig.6a) there is a rapid decay of the
empirical autocorrelation function at the time of the order of one
day followed by a more slow decrease (solid line). The generated
curves reproduces this abrupt drop (dashed line) at time of the
order of 1.5 days. The same behavior is also exhibited for S\&P500
Index \cite{F03}. The periodic oscillations of the empirical
correlation function arises from a stable increase of the trade
activity at both the  beginning and the end of day.

In the case of low-frequency dynamics the generated
autocorrelation function reproduces the initial drop of the
empirical curve at times of the order of two months (Fig.6b).
\section*{\large 6. Conclusion}
The SV models introduced in \cite{S87,SS91,H93,HW87} define the
volatility as a Markovian process. On the other hand the
volatility autocorrelation function shows the existence of two or
more characteristic times,which, in general, is not typical for
the Markovian processes. In recent works \cite{B01,F03,F3} the SV
models have been considered, describing  volatility as a
superposition of Markovian processes with  different
characteristic times. Using these approaches we consider
volatility at both short and long time scales in the Markovian
approximation. On the basis of the empirical data, employing
Eqs.(\ref{10}) and (\ref{11}), we estimated the coefficients
$D_{1,2}$ of Ito SDE defining the volatility dynamics. It has been
shown that for the long time scale the empirical data support the
ExpOU model with the characteristic time $\alpha^{-1}=2.43$ months
(see Eqs.(\ref{16}) and (\ref{17})). On the short time scale the
drift coefficient can also be described within the scope of this
model with $\alpha^{-1}=0.79$ days (Eq.(\ref{14})). As regards the
diffusion coefficient it shows more complicated behavior than a
simple square dependence (see Eq.(\ref{15})).

On the base of the numerical solution of the Fokker-Plank equation
we have considered the time evolution of the conditional PDF
$p(\sigma,t|\sigma_0,t_0)$. It has been shown that the stationary
state settles in accordance with the found relaxation times within
time of the order $t\sim 2/\alpha$, where for the long time scale
$t_L\sim 5$ months and for short time scale $t_S\sim 1,5$ days.
Very good agreement is found between the calculated stationary
densities and the empirical curves (Fig.4). This fact supports the
validity of estimating of the coefficients $D_{1,2}$.

On the basis of Eqs (\ref{20}) and (\ref{21}) the price series
$x_t$ has been generated and PDFs of the price changes $\Delta x_t
= x_t - x_{t-\Delta}$ for the different time delays $\Delta$
obtained. In the absence of a correlation between the Wiener
processes $W_1(t)$ and $W_2(t)$ the agreement between the
simulated and empirical densities  for both time scales is good
enough (Fig.5).

On the basis of the generated data the volatility autocorrelation
function has been obtained. The empirical autocorrelation function
within  the interval of a few months shows the existence of at
least two characteristic  time scales (Fig.6). At the initial
stage there is a drop within approximately 0.5 days followed by a
slow decay. As it is seen from Fig.6  the generated data  at both
short (within $\sim 0.5$ days) and long (within $\sim 2$ months)
time scale separately describe  such behavior.

The obtained results  deserve attention especially if one takes
into account that the parameters $D_{1,2}$ were not fitted
specially to neither the empirical densities  nor the behavior of
autocorrelation function.

The parameters of the ExpOU model for the Dow-Jones Index  were
estimated in other works also. In \cite{PM05} it has been shown
that  the fitting of the theoretical curve  to the empirical
volatility autocorrelation function at the  interval of a few
months gives the estimation of the relaxation time of the order of
$35\pm 18$ days. On the upper bound this estimation is close to
our results for LFD.

On the other hand the derivation of SDE for the volatility  on the
basis of the empirical data was also considered in \cite{WO98}.
The equation obtained in this work  for the low-frequency dynamics
of the Dow-Jones Index  is close enough to the Eq.(\ref{4}) of the
ExpOU model. Numerical solution the the Fokker-Plank equation for
the conditional PDF has shown that the stationary state settles
within 3-4 months, which approximately corresponds to our data.

Thus the results reported in this work show that the employment of
the Markovian approximation at the individual time scale in all
probability allows to describe at this scale a number of market
appropriateness. In particular we have shown that the volatility
autocorrelation function and the probability return densities
obtained within the scope of this approximation  are consistent
enough with their empirical analogues separately at both short and
long time scales.

\newpage
\section*{Figures}

\begin{figure}[htb]
  \centering
 \includegraphics[width=15.8cm,height=14.4cm]{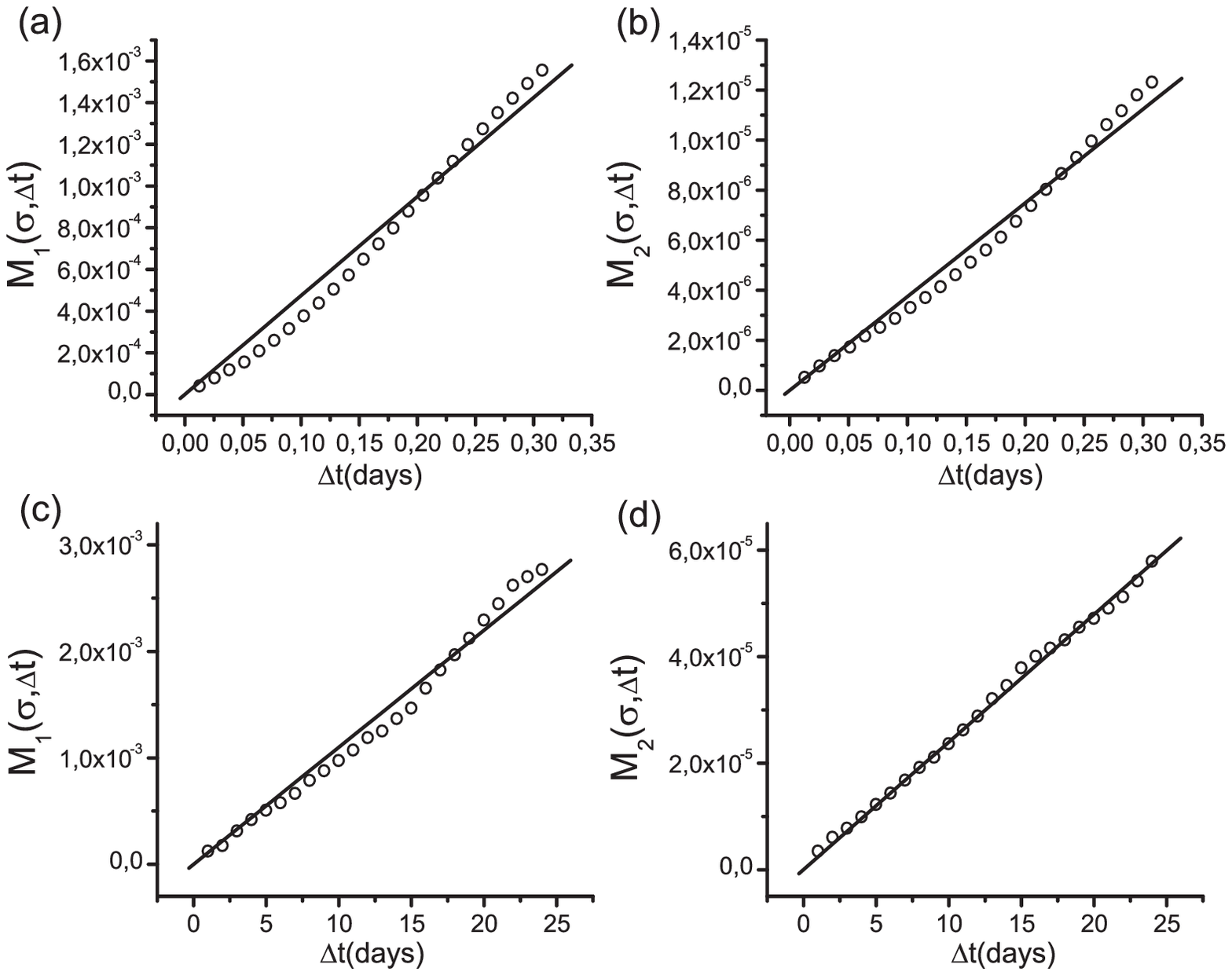}
 \caption{The typical dependence of $M_{1,2}(\sigma , \Delta
t)$ on $\Delta t$ at  small $\sigma$ (solid line). Circles
correspond to the quantities obtained from Eq.(\ref{11}).
Figs.1a,1b correspond to HFD set at $\sigma = 0.032$; Figs. 1c, 1d
correspond to LFD set at $\sigma = 0.02$.}
\end{figure}

\begin{figure}[htbp]
  \centering
 \includegraphics[width=15.8cm,height=14cm]{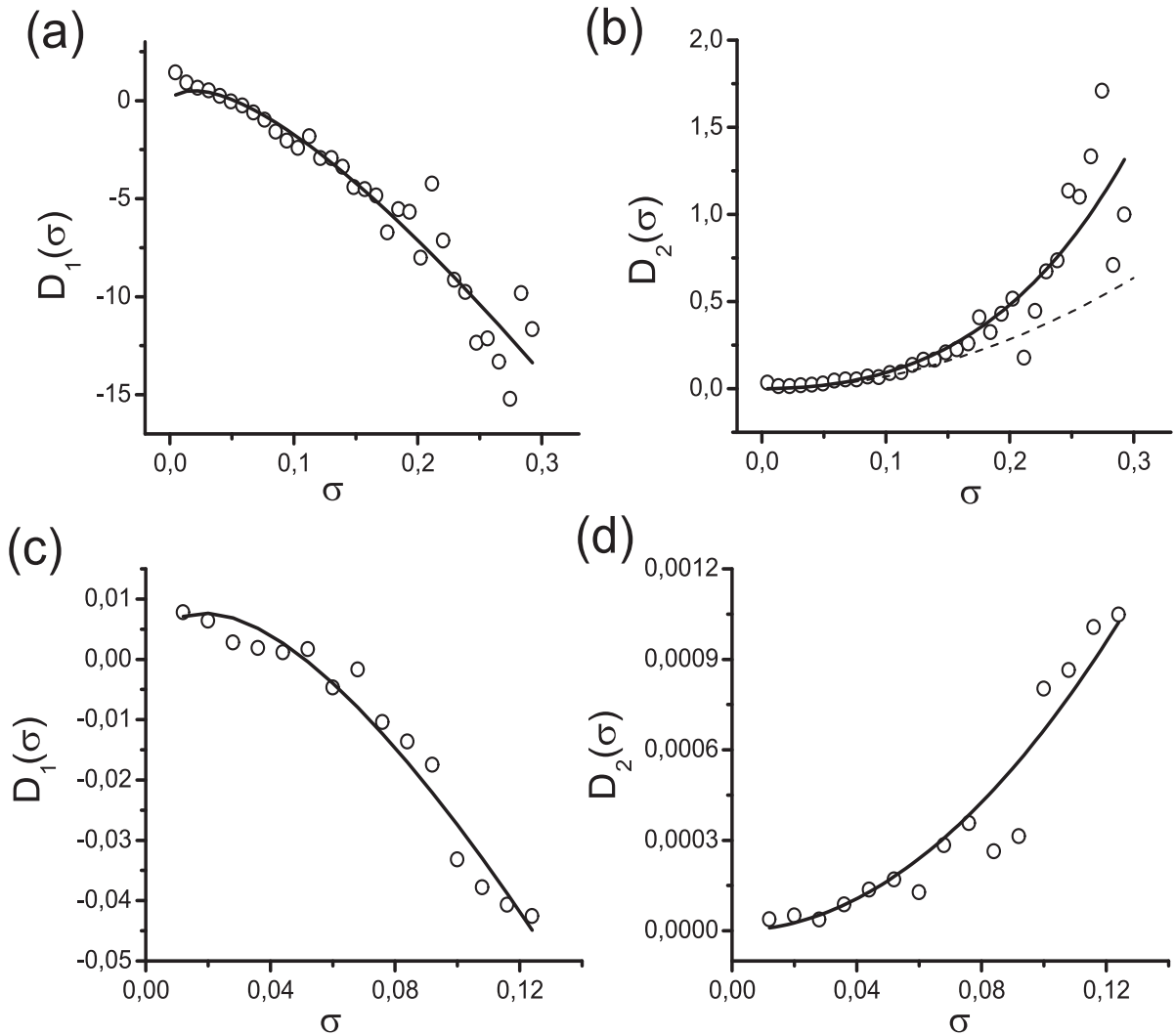}
 \caption{The coefficients $D_{1,2}(\sigma , t)$
obtained from the empirical data using Eqs. (\ref{11}) and
(\ref{12}) (circles). The solid line is the approximation by the
smooth curve. Figs.2a and 2b correspond to HFD set;  Figs.2c and
2d correspond to LFD set. The dashed line in Fig. 2b shows the
square dependence of $D_2$ on $\sigma$.}
\end{figure}

\begin{figure}[htbp]
  \centering
 \includegraphics[width=15.2cm,height=7cm]{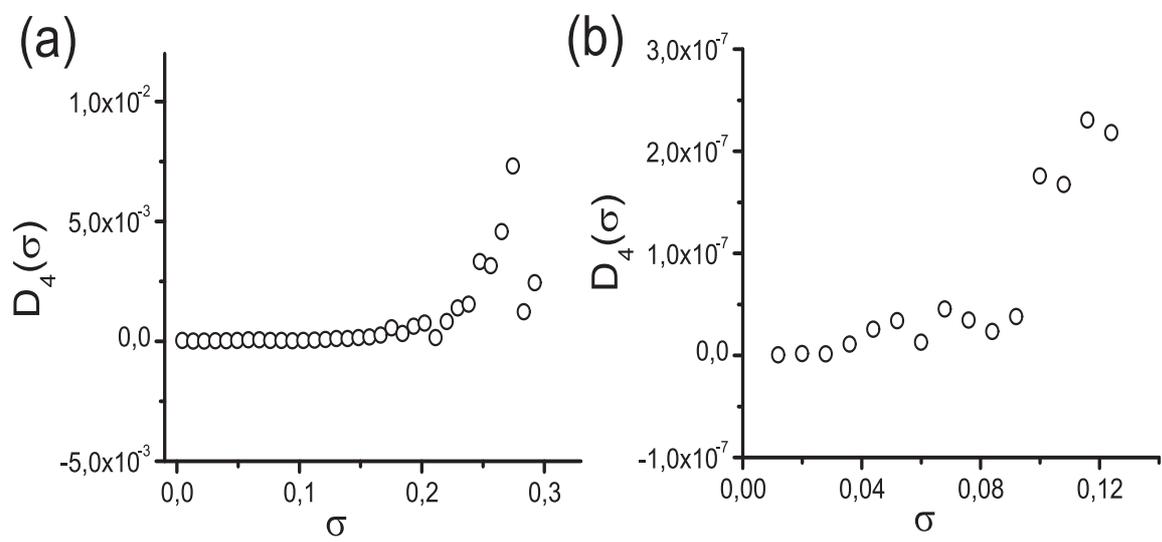}
    \caption{The dependence of the coefficient the $D_4$
on $\sigma$ ; (a) - HFD set; (b) - LFD set.}
\end{figure}

\begin{figure}[htbp]
  \centering
 \includegraphics[width=14.8cm,height=18cm]{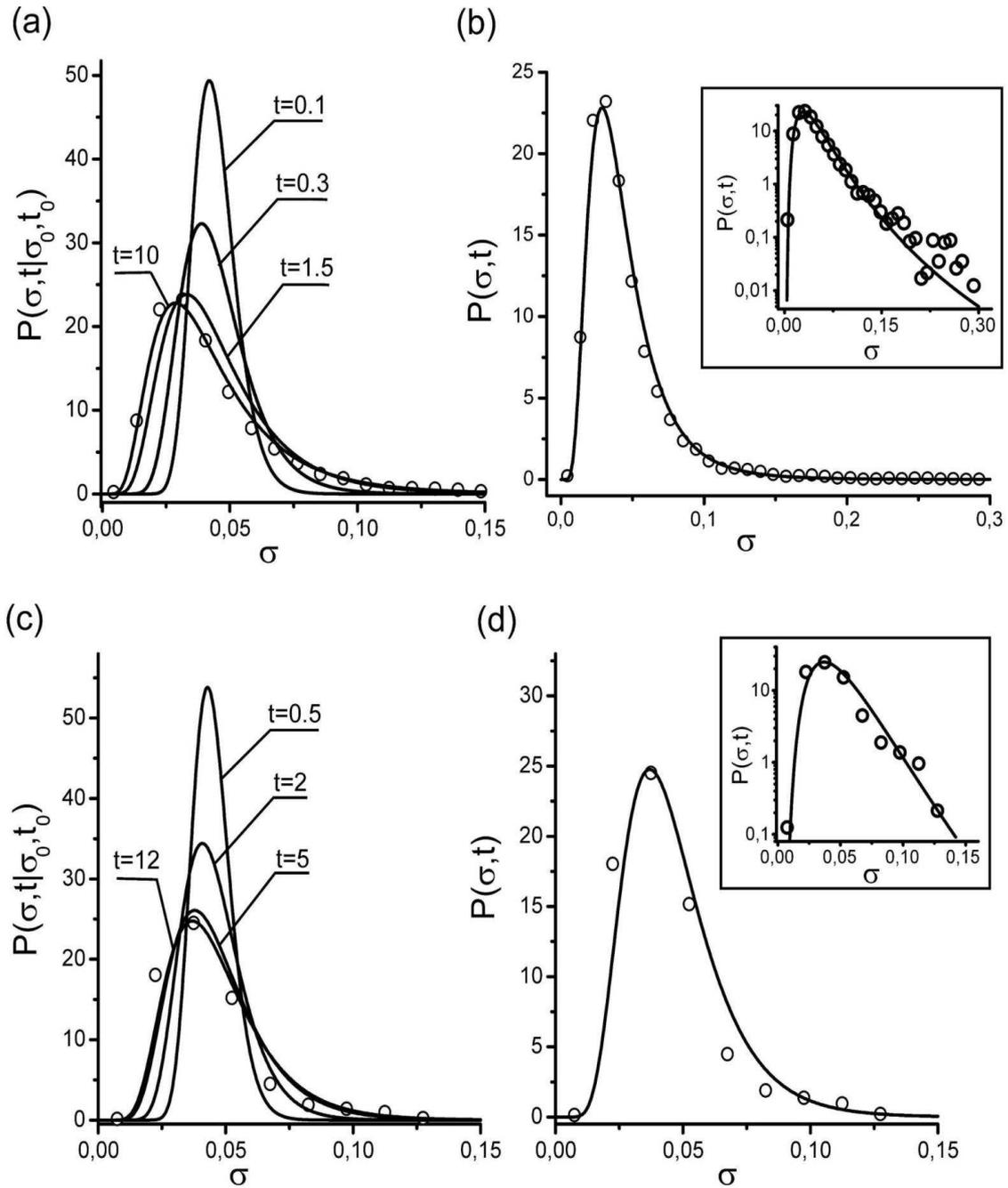}
 \caption{The time evolution of the conditional PDFs
$p(\sigma t|\sigma_0 t_0)$; (a) - HFD set, $t = 0.1; 1.5; 10; 12$
days; (c) - LFD set,  $t = 0.5; 2; 5; 12$ months; $t_0 = 0$. The
solid line in Figs. 4b and 4d correspond the stationary
distributions obtained from the Fokker-Plank equation,
respectively, for $t = 10$ days (HFD) and $t = 12$ months (LFD).
Circles - the empirical distributions. The inset shows the same
curves in the log - log scale.}
\end{figure}

\begin{figure}[htbp]
  \centering
 \includegraphics[width=15.8cm,height=8cm]{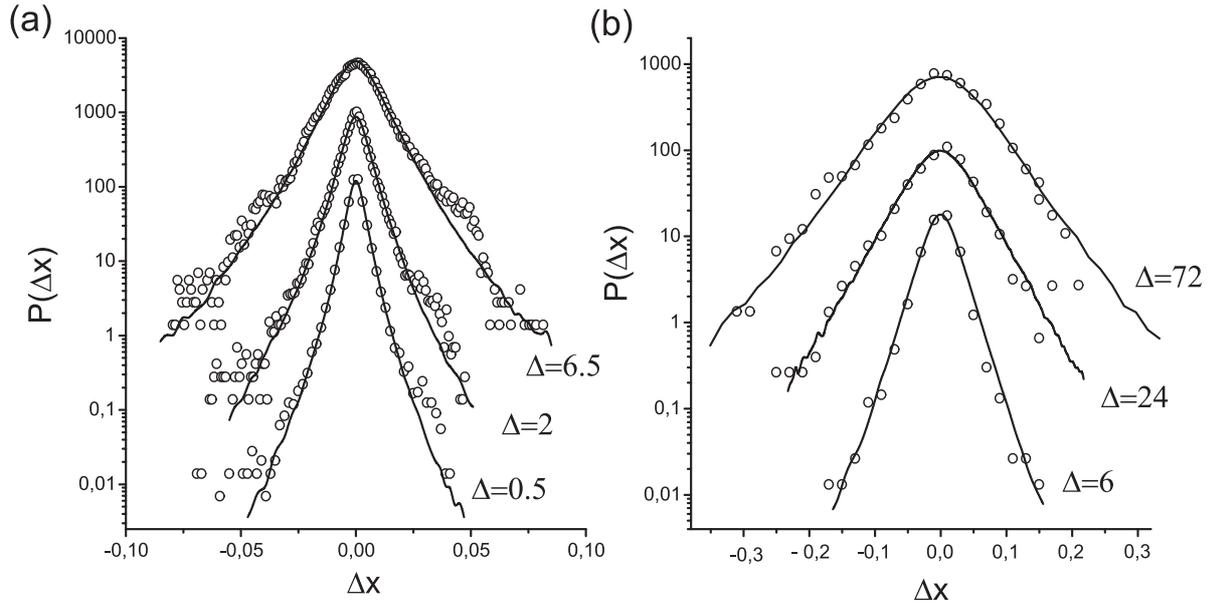}
 \caption{The probability densities of the price
changes $\Delta x = \ln S_t/S_{t-\Delta} - \mu \Delta$. The solid
line  have been obtained  from the simulated data; circles - the
empirical data;(a) - HFD set, $\Delta = 0.5; 2; 6.5$ hours; (b) -
LFD set, $\Delta = 6; 24; 72$ days. For convenience of
presentation the PDFs are shifted in vertical direction by
multiplication by 10.}
\end{figure}

\begin{figure}[htbp]
  \centering
 \includegraphics[width=14.1cm,height=6.8cm]{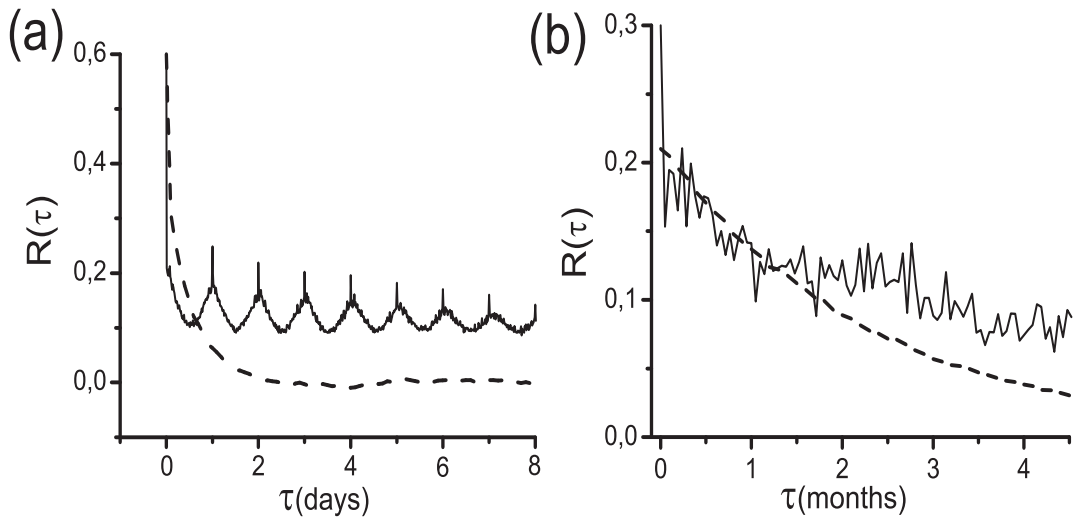}
 \caption{The autocorrelation function of $|r_t|$.
The dashed line have been obtained from the generated data, the
solid line obtained on the basis the empirical data; (a) - HFD
set, (b) - LFD set.}
\end{figure}

\end{document}